\documentclass[aps,prl,twocolumn,groupedaddress,showpacs,amsmath,preprintnumbers]{revtex4}
\input{amssym}
\usepackage{graphicx}
\usepackage{hyperref}
\usepackage{dcolumn}% Align table columns on decimal point
\usepackage{bm}

\def\be {\begin{equation}}

\def\ee  {\end{equation}}

\def\bea {\begin{eqnarray}}

\def\eea {\end{eqnarray}}

\begin{document}
\preprint{gr-qc/04}
\title[]{The Spectrum of Fluctuations in Singularity-free Inflationary Quantum Cosmology}
\author{Stefan Hofmann$^{1}$ and Oliver Winkler$^{1,2}$}
\email[]{shofmann@perimeterinstitute.ca, oliver@math.unb.ca}
\affiliation{$^{1}$Perimeter Institute for Theoretical Physics,
Waterloo, ON,
Canada, N2L 2W9\\
$^{2}$ Department of Mathematics and Statistics, University of New
Brunswick, Fredericton, NB, Canada E3B 5A3} \pacs{98.70.Vc,
98.80.Cq, 98.80.Qc}
\date{\today}
\begin{abstract}
We calculate the power spectrum of vacuum fluctuations of a generic
scalar field in a quantum cosmological setting that is manifestly
singularity-free. The power spectrum is given in terms of the usual
scale invariant spectrum plus scale dependent corrections. These are
induced by well-defined quantum fluctuations of de Sitter spacetime,
resulting in a modified dispersion relation for the scalar field.
The leading correction turns out to be proportional to the ratio of
Hubble scale and Planck Mass. The maximal relative change in the
spectrum is on the ten percent level and might be observable with
future CMB experiments.
\end{abstract}

%  JCAP uses keywords rather than pacs numbers.

\begin{flushleft} {\bf Keywords}:
\end{flushleft}


\maketitle

%\section{Introduction}
A fundamental description of the physics during the Planck epoch
remains elusive. Important issues like the big bang singularity,
the unknown laws that govern the dynamics of all physical degrees
of freedom during that period and the correct initial conditions
are not yet fully understood.

The success of inflationary cosmology in explaining an
overwhelming amount of data of unprecedented precision can be
considered as a proof that the cosmological evolution suppresses
to a large extent imprints from the Planck epoch. Nevertheless, at
least in principal, some cosmological observables might have
retained a memory of initial conditions and dynamics of the Planck
epoch. Especially the cosmic microwave background (CMB) radiation
should bear imprints from the Planck epoch, through its dependence
on the spectrum of inflationary density fluctuations.

In recent years, a growing body of researchers have investigated
the impact of fundamental (or unknown) physics through this
cosmological window \cite{mar:01} and the new field is often
referred to as {\it
transplanckian physics} \cite{bra:00}.
An important issue addressed there is the
magnitude with which processes of characteristic energies around a
fundamental scale $M_{\rm P}$ (e.g.~the string or the Planck
scale) couple to known physics far below $M_{\rm P}$. On
dimensional grounds this coupling is expected to be proportional
to $(H/M_{\rm P})^r$, where $H$ denotes the inflationary Hubble
scale. The power $r$ depends on the details of the specific
transplanckian model that has been chosen. Even if the
inflationary Hubble scale is at the GUT scale, $r\le 1$ is
required to extract Planck imprints from the CMB, in order to
disentangle them from uncertainties due to cosmic variance.

Instead of modelling physics at the Planck scale, the effective
field theory approach \cite{sch:04,col:05}  allows a more
systematic study of high-energy imprints. Physical processes
characterized by energy scales above $M_{\rm P}$ can be related to
the dynamics below $M_{\rm P}$ by only a finite number of
couplings. However, the Wilson renormalization group approach
presumes a Lagrange description, or an action. To pose a
well-defined initial value problem, boundary conditions have to be
specified that mimic the corresponding initial conditions in the
Hamilton description of the field theory. This is the {\it
Boundary Effective Field Theory} (BEFT) formalism. Instead of
setting the boundary conditions for all modes on an equal time
hypersurface, boundary conditions might be set when the physical
momentum of a mode is redshifted to a physical cut-off scale. This
boundary proposal is called the {\it New Physics Hypersurface}
(NPH) formalism \cite{dan:02}. Both, the BEFT and the NPH proposal
allow for $r=1$ modifications of the inflationary power spectrum,
see \cite{gre:05,eas:05} and  \cite{lbe:02}. Both boundary
proposals modify the initial state of the quantum fluctuation that
grow into cosmological perturbations. However, the authors of
\cite{kal:02} find an effective action based on low energy
locality allowing only for $r=2$ corrections to the power
spectrum. They claim that the irrelevant operators added to
Einstein gravity in the BEFT approach represent a boundary
condition that presumably violates low energy locality.

Here, by contrast, we derive a modified dispersion relation from
an underlying quantum theory of de Sitter spacetime, concrete
models are given in \cite{ash:03,win:04}. Their characteristic
feature is a careful construction of the inverse scale operator
$a^{-1}$, with finite eigenvalues even at the classical
singularity. While these eigenvalues approach the expected value
$1/a_{\rm cl}$ in the large scale limit, they also carry quantum
corrections which lead to modified evolution equations for the
minimally coupled scalar field. We take these corrections into
account by replacing the classical quantities $a_{\rm cl}$ and
$a_{\rm cl}^{\; -1}$ by the expectation values of the
corresponding operators in the evolution equations. This
corresponds to a mean field approximation for the gravitational
degree of freedom.

In this framework, we derive the evolution equations for the
vacuum fluctuations in a general form, valid for any quantum
cosmology that is singularity-free in the sense of a bounded
inverse scale operator. For the prediction of the fluctuation
spectrum we then use the concrete model discussed in
\cite{win:04}.

%\section{General evolution equations}
The quantum theory of the scalar field in the canonical approach
is formulated in the dynamical variables $(\Phi, \pi)$ on a de
Sitter background. The Hilbert space for the coupled
gravity-scalar field system is $\cal H = {\cal H}_{\rm g} \otimes
{\cal H}_{\rm s}$ where ${\cal H}_{\rm g}$ denotes the geometrical
Hilbert space and ${\cal H}_{\rm s}$ is the Hilbert space of the
scalar field. As discussed above, the quantum dynamics of the
combined system is reduced to dynamics for the scalar field
operator on a de Sitter background including quantum corrections
in the mean field approximation. The mean field approximation is
characterized by expectation values $\langle a\rangle$ and
$\langle a^{-1}\rangle$ of the operators corresponding to the
scale factor and the inverse scale factor with respect to
appropriate quantum cosmological states. As we aim at deriving the
effective evolution equation in as general a form as possible, we
will leave this state unspecified until we specialize to the
concrete framework of \cite{win:04}.

The effective Hamilton operator for the scalar field theory on this
background is then given by
\begin{equation}
\langle H_{\rm s}\rangle = \int \langle {\rm vol} \; \rho_{\rm s} \rangle
\; ,
\end{equation}
with the energy density
\begin{equation}
\label{ed} \langle \rho_{\rm s}\rangle = \frac{1}{2} \left(\langle
a^{-3}\rangle \pi\right)^2 + \frac{1}{2} \left(\langle a^{-1}\rangle
\nabla \Phi \right)^2 + V(\Phi) \; .
\end{equation}

The dynamics of the field operators on the FRW background is given
by Heisenberg's equations of motion
\begin{eqnarray}
\dot{\Phi} &=& \langle a^3\rangle \left(\langle a^{-3}\rangle
\right)^2 \pi
\; ,\\
\dot{\pi} &=& \langle a^3\rangle \left[ \left(\langle a^{-1}\rangle
\nabla\right)^2\Phi - \frac{{\rm d}V}{{\rm d}\Phi} \right] \; ,
\end{eqnarray}
where dots denote derivatives with respect to cosmic time.
Heisenberg's equations are equivalent to the second order
differential equation for the scalar field
\bea
\label{eom1} \ddot{\Phi} &=& \left[ 3 \frac{\dot{\langle
a\rangle}}{\langle a\rangle} + 2 \frac{\dot{\langle
a^{-3}\rangle}}{\langle a^{-3}\rangle} \right]
\dot{\Phi} \nonumber \\
&& + \left(\langle a^3\rangle \langle a^{-3}\rangle\right)^2 \left[
\left(\langle a^{-1}\rangle \nabla\right)^2 \Phi - \frac{{\rm
d}V}{{\rm d}\Phi} \right]  \; .
\eea

Following the standard procedure, we expand the scalar field
around its homogeneous expectation value with respect to an
arbitrary but fixed vacuum state and Fourier transform the quantum
fluctuations around the homogeneous state. The fluctuations are
conveniently expressed in terms of time dependent oscillators
$A(k,t)$ and $A^\dagger(-k,t)$, where $k$ denotes the comoving
wavenumber characterizing the fluctuation. The initial conditions
imposed on these solutions correspond to a choice of the vacuum
state $|\Omega, t_{\rm i}\rangle$ at the initial time $t_{\rm i}$.
$A(k,t_{\rm i})$ annihilates $|\Omega, t_{\rm i}\rangle$. The
vacuum state defined this way is often referred to as the lowest
energy state, the minimal uncertainty state with respect to
$\Delta \Phi \Delta \pi$ or the instantaneous Minkowski vacuum.
For $t_{\rm i} \rightarrow -\infty$ this vacuum proposal includes
the Bunch-Davies vacuum state.

The time evolution given by Bogolubov transformations
mixes annihilation and creation operators.
In terms of annihilation and creation operators the
fluctuations are given by
\begin{equation}
\Phi(k,t) = \phi(k,t) A(k,t_{\rm i})
+ \phi^*(k,t) A^\dagger(k,t_{\rm i})
\; ,
\end{equation}
with $\phi$ denoting the corresponding mode functions,
satisfying the reality condition $\phi(k,t)=\phi^*(-k,t)$.

Close to the classical de Sitter limit we can expand the
expectation values in the gravitational sector around their
classical values: $\langle a\rangle = a_{\rm cl}$ and $\langle
a^{-1}\rangle = a_{\rm cl}^{\; -1} (1+\langle a^{-1}\rangle_q)$,
where $q$ denotes the quantum corrections to de Sitter spacetime
and $\langle a^{-1}\rangle_q \ll 1$. In the quantum cosmological
model developed in \cite{win:04}
, $\langle a^{-1}\rangle_q =
\sqrt{2\pi} a_{\rm cl}^{-1} + {\cal O}(1/a_{\rm cl}^{\; 2})$.

Let us introduce the dimensionless variable $x=a/a_{\rm i}$, with
$a_{\rm i}\equiv a(t_{\rm i})$, the value of $a$ at a fixed
initial time $t_i$. In the quantum cosmological models
\cite{ash:03,win:04} the initial scale factor can be fixed very
close to the Planck scale, $a_{\rm i} = \beta \sqrt{8\pi} L_{\rm
p}$, with $\beta$ being a parameter constrained by the requirement
of having a consistent perturbation analysis. We find $\beta={\cal
O}(1-10)$, since the quantum corrected spectrum of the inverse
scale factor approaches the classical value very fast for
$a>\sqrt{8\pi} L_{\rm P}$. As we are interested in the
semiclassical regime only, we focus on the case $x>1$. Then, the
fluctuation modes $\phi(k,x)$ obey
\begin{eqnarray}
\label{eomx}
\frac{{\rm d}^2\phi}{{\rm d}x^2}
&=&
-\left(4+3 f_{H}(x)\right) \frac{1}{x}
\frac{{\rm d}\phi}{{\rm d}x}
\nonumber \\
&&-\left(1+f_k(x)\right) \frac{1}{x^4} \frac{k/a_{\rm i}}{H} \phi
\; ,
\end{eqnarray}
with the leading quantum correction
$f_H(x) \equiv (\sqrt{8\pi} L_{\rm P}/a_{\rm i}) 1/x$ to the Hubble friction
and $f_k(x) \equiv 4 f_H(x)$ to the redshift term.

\begin{figure}[t]
\vspace{-5pt}
\includegraphics[width=6cm]{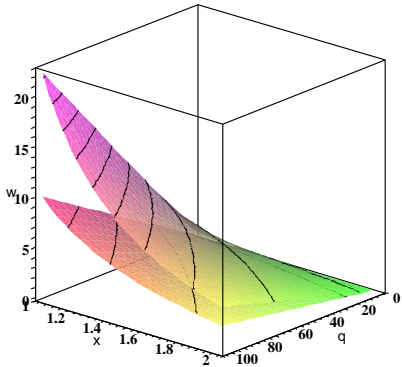}
\vspace{-5pt}
\caption{The dispersion relation on classical de Sitter
(bottom) and on quantum corrected de Sitter spacetime (top).
Contours are curves of constant $\omega$.
Units on the vertical axis are arbitrary}
\label{fig1}
\end{figure}

Using $\phi \rightarrow x^{-2} \exp{(3 f_H(x)/2)} \phi$,
(\ref{eomx}) can be transformed into
\begin{equation}
\left[\frac{{\rm d}^2}{{\rm d}x^2} + \omega^2(x)\right] \phi(k,x) = 0
\; ,
\end{equation}
with $\omega^2 \equiv \omega_{\rm cl}^{\; 2} + \omega_{\rm q}^{\;
2}$, see \ref{fig1}.
Here, $\omega_{\rm cl}^{\; 2} \equiv (1/x)^4 (k/a_{\rm
i}/H)^2 - 2/x^2$ is the frequency of an oscillator on classical de
Sitter spacetime. The time dependence in the classical dispersion
relation is due to the space expansion, causing redshift and
damping of fluctuations. The leading quantum corrections to de
Sitter spacetime modify the dispersion of the fluctuations:
\begin{eqnarray}
\omega_{\rm q}^{\; 2}
&\equiv&
\frac{1}{x^2}\left[
\frac{1}{x^2} \left(\frac{k/a_{\rm i}}{H}\right)^2 f_k(x)
\right. \nonumber \\
&& - \frac{3}{2} \left.
\left(3  + x \frac{{\rm d}}{\rm dx} + \frac{3}{2}\right) f_H(x)
\right]
\; .
\end{eqnarray}
Deep inside the Hubble radius ($k/aH \gg 1$), the redshift term
dominates the dispersion relation. In this regime, the leading modification
of the dispersion relation decays like $1/\sqrt{x}$ relative to the classical
redshift. In the super-Hubble ($k/aH \ll 1$) regime, the friction term generated
by the Hubble expansion dominates and the leading modification has the same asymptotic
behavior like the redshift correction for $x\gg1$.

The dynamics of fluctuation modes on subhorizon scales is altered
mainly due to the redshift modification induced by quantum corrections
of de Sitter spacetime. More precisely, the redshift correction modifies
the oscillation frequency, while it has only minor influence on the oscillation
amplitude. The amplitude is influenced by quantum corrections to
the friction term induced by the Hubble expansion. The asymptotics
of modes on sub-Hubble scales can be extracted for $x\gg 1$ from
\begin{eqnarray}
\label{sub}
\phi_{\rm sub}(x)
&=&
\frac{1}{a(2k)^{1/2}} \frac{\exp{(3 f_H(x)/2)}}{(1+f_k(x))^{1/4}}
\nonumber \\
&&
\times\exp{\left({\rm i} \frac{k/a_{\rm i}}{H} \frac{(1+f_k(x))^{3/2}}{x f_k(x)}\right)}
\; .
\end{eqnarray}
The integration constants have been determined by the requirement that
on sub-Hubble scales (but still on energy scales below $M_{\rm P}$),
the vacuum is populated by quantum fluctuations with positive
frequencies only, normalized as in the standard theory of scalar fields
on Minkowski spacetime.

\begin{figure}[t]
\vspace{-5pt}
\includegraphics[width=6cm]{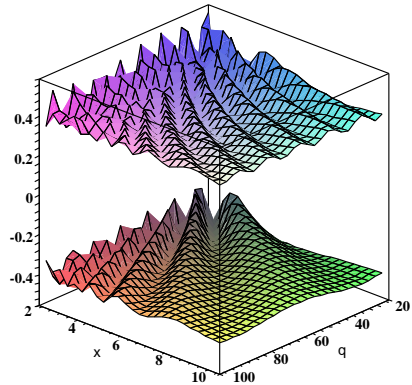}
\vspace{-5pt}
\caption{The real part of $\phi_{\rm sub}$ on classical de Sitter (bottom)
and on quantum corrected de Sitter spacetime (top). Both cases have been separated
by adding/substracting a constant. Units on the vertical axis are arbitrary }
\label{fig2}
\end{figure}

Fig.\ref{fig2} shows the subhorizon evolution of vacuum fluctuations on the standard
de Sitter and on the quantum corrected de Sitter spacetime. It can be seen
that the main effect of the quantum corrected spacetime is to roughly quadruple
the oscillation frequencies. As usual, the oscillation decays
like $1\sqrt{k}$ for constant $x$ and increasing wavenumber,
and decreases like $1/x$ for $x\gg 1$ and constant $k$.

Vacuum fluctuations on super-Hubble scales are mostly effected by quantum corrections
to the Hubble expansion. However, there is a mild dependence on the redshift corrections
through the matching of the dynamics on sub- and super-Hubble scales
at horizon crossing. The asymptotic
behavior of the vacuum fluctuations for $x\gg 1$ can be extracted from
\begin{eqnarray}
\phi_{\rm sup}(x)
&=&
\frac{1}{a\sqrt{2k}} \frac{\exp{\left(3 f_H(x)/2\right)}}{(1+f_k(x_{\rm c}))^{1/4}}
\nonumber \\
&&\times \frac{{\rm W_M}(-1,3/2,3f_H(x))}{{\rm W_M}(-1,3/2,3f_H(x_{\rm c}))}
\; .
\end{eqnarray}
Here, ${\rm W_M}$ denotes the Whittaker function that is related
to Kummer's function $M$, see \cite{AS}. $x_{\rm c}\equiv
k/(a_{\rm i}H)$ is the value of the scale factor when the vacuum
fluctuation characterized by comoving wavenumber $k$ crossed the
Hubble radius, normalized to the initial scale factor. The dependence on
$x_{\rm c}$ is chosen in order to match the solutions on sub-Hubble scales
(\ref{sub}) at horizon crossing. In this way the choice of a
vacuum proposal determines the otherwise unknown integration
constants in the super-Hubble solutions. On super-Hubble scales,
the general solution is actually a linear combination of both
Whittaker functions ${\rm W_M}$ and ${\rm W_W}$. For $x\gg 1$,
${\rm W_M}\propto 1/x^2$ while ${\rm W_W}\propto x$. Hence, only
${\rm W_M}$ has the correct asymptotic behavior.

We are now ready to present the main result of this letter: the linear power spectrum
(defined as ${\cal P}_{\Phi} \equiv (k^3/2\pi^2) \langle \Omega|(\Phi^\dagger \Phi)(k,x_{\rm c})|\Omega\rangle$)
of quantum fluctuations of a generic scalar field around its vacuum state:
\begin{eqnarray}
{\cal P}_{\Phi} (k, x_{\rm c})
&=&
\left(\frac{H}{2\pi}\right)^2 \Bigg[1+3f_H(x_{\rm c}) -\frac{1}{2} f_k(x_{\rm c})
\nonumber \\
&&\hspace{1.5cm} + {\cal O}\left[(H/M_{\rm P})^{2}\right]\Bigg] \;
.
\end{eqnarray}
The leading quantum correction to the Hubble friction at horizon crossing
is given by $f_H(x_{\rm c}) = f_H(1)/x_{\rm c}$, with
$f_H(1)\equiv \sqrt{8\pi} L_{\rm P}/a_{\rm i}$ and
$f_k(x_{\rm c}) = 4 f_H(x_{\rm c})$.

Note that quantum corrections to the redshift evolution lower the statistical power
of large wavelength modes, while corrections of the Hubble friction
enhance the statistical power on large scales.
However, both the quantum correction to the Hubble friction and the correction
to the cosmological redshift decay like $1/x_{\rm c}$. So vacuum fluctuations
characterized by large wavenumbers have less statistical support compared to modes
with smaller wavenumbers.

We finally find for the linear power spectrum
\begin{eqnarray}
\label{res}
{\cal P}_{\Phi} (k,x_{\rm c})
&=&
\left(\frac{H}{2\pi}\right)^2
\Bigg[1+f_H(1) \left(\frac{k/a_{\rm i}}{H}\right)^{-1}
\nonumber \\
&&\hspace{1.5cm} +
{\cal{O}}\left[\left(H/M_{\rm P}\right)^2\right] \Bigg]
\; .
\end{eqnarray}

In order to estimate the size of the leading modification
to the classical power spectrum ${\cal P}_\Phi^{\rm cl}$,
we note that $\delta {\cal P}_\Phi / {\cal P}_\Phi^{\rm cl} \equiv
{\cal P}_\Phi/{\cal P}_\Phi^{\; \rm cl} - 1 =
(\sqrt{8\pi} L_{\rm P}/a_{\rm i}) (1/x_{\rm c}(k))$.
Let us estimate $a_{\rm c}$ relative to the initial scale factor
$a_{\rm i}$ for the CMB quadrupole:
\begin{equation}
\label{est}
\frac{1}{x_{\rm c}}\Bigg|_{l=2}
=
\frac{1}{x_0} \frac{H}{H_{\rm 0}}
\; .
\end{equation}
Here, $H$ and $H_0$ denote the Hubble radius during inflation and
today, respectively. Now, $x_0= (a_0/a_{\rm e}) (a_{\rm e}/a_{\rm
i})$, with $a_{\rm e}$ denoting the scale factor at the end of
inflation. Assuming instantaneous reheating, $a_0/a_{\rm e}
\approx T_{\rm rh}/T_0 \approx \sqrt{H M_{\rm P}}/T_0$, where
$T_0$ is the current temperature of the CMB photons and $T_{\rm
rh}$ is the reheating temperature. We parameterize the ratio of
the scale factor at the end of inflation and the initial scale
factor by the number of e-folds before the end of inflation,
$a_{\rm e}/a_{\rm i}\equiv \exp{(N)}$. Then (\ref{est}) becomes
\begin{eqnarray}
\frac{1}{x_{\rm c}}\Bigg|_{l=2}
&\approx&
\exp{(-N)} \frac{T_0}{H_0} \sqrt{\frac{H}{M_{\rm P}}}
\nonumber \\
&\approx&
10^{29} \exp{(-N)} \sqrt{\frac{H}{M_{\rm P}}}
\; .
\end{eqnarray}
For the scale corresponding to the CMB quadrupole we find
$x_{\rm c} = 1$ for $N\approx 62$, assuming $H/M_{\rm P} \approx 10^{-4}$.

As mentioned earlier, in the quantum cosmological model we use here, \cite{win:04},
the initial scale factor can be fixed close to the Planck scale,
$a_{rm i} = {\cal O}(1-10) \sqrt{8\pi} L_{\rm P}$.
For the scale corresponding to the CMB quadrupole we therefore find
\begin{equation}
\frac{\delta {\cal P}_\Phi}{{\cal P}_\Phi^{\rm cl}}\Bigg|_{l=2}
\approx
{\cal O}(10) \%
\; .
\end{equation}

In summary, we have derived the modified dispersion relation
for a generic scalar field on de Sitter space subject to
quantum gravitational fluctuations in the framework of
a singularity-free quantum cosmology. We have calculated the corresponding corrections to
the power spectrum and find the leading correction term to be of order $H/M_P$,
which would be promising for future CMB experiments.
For the quadrupole scale, we expect an ${\cal O}(10)\%$ modification of the classical
power spectrum.

It is a pleasure to thank R. Brandenberger and V.
Husain for helpful discussions.

\end{document}